\documentclass[twocolumn,article]{jjap2} 
\title{Terahertz radiation by optical rectification in a hydrogen-bonded organic molecular ferroelectric crystal, 2-phenylmalondialdehyde}

\author{Wenguang \textsc{Guan}$^{1}$, Noriaki \textsc{Kida}$^{1}$\thanks{To whom correspondence should be addressed. E-mail: kida@k.u-tokyo.ac.jp}, Masato \textsc{Sotome}$^{1}$, Yuto \textsc{Kinoshita}$^{1}$, Ryotaro \textsc{Takeda}$^{1}$, Akito \textsc{Inoue}$^{2}$, Sachio \textsc{Horiuchi}$^{3,4}$ and Hiroshi \textsc{Okamoto}$^{1}$}

\inst{$^{1}$Department of Advanced Materials Science, The University of Tokyo, Kashiwa, Chiba 277-8561, Japan \\
$^{2}$Department of Applied Physics, The University of Tokyo, Bunkyo, Tokyo 113-8656, Japan \\
$^{3}$National Institute of Advanced Industrial Science and Technology (AIST), Tsukuba, Ibaraki, 305-8562, Japan \\
$^{4}$CREST, Japan Science and Technology Agency (JST), Chiyoda, Tokyo 102-0076, Japan}

\abst{Terahertz radiation by optical rectification has been observed at room temperature in a hydrogen-bonded organic molecular ferroelectric crystal, 2-phenyl malondialdehyde (PhMDA). The radiated electromagnetic wave consisted of a single-cycle terahertz pulse with a temporal width of $\sim$ 0.5 ps. The terahertz radiation amplitude divided by the sample thickness in PhMDA was nearly equivalent to that in a typical terahertz wave emitter ZnTe. This is attributable to a long coherence length in the range of 130 $\sim$ 800 $\mu$m for the terahertz radiation from PhMDA. We also discussed the possibility of PhMDA as a terahertz wave emitter in terms of the phase-matching condition.}

\kword{Organic ferroelectrics, Terahertz radiation, Nonlinear optical process, Ferroelectric domains, Imaging}

\begin{document}
\maketitle

\section{Introduction}

Ferroelectrics are versatile compounds used in various electronic and photonic devices [1]. They are partly used as constitute elements of actuators owing to their large piezoelectric effect [2,3]. Another important characteristic of ferroelectrics is their large nonlinear optical effect [1]. For example, LiNbO$_3$ is used as the wavelength converter of light via the nonlinear optical effect [4,5], i.e., optical parametric oscillation (OPO) and emission of terahertz radiation by difference frequency generation (DFG). In these applications, inorganic ferroelectrics are mainly used. Low-dimensional organic molecular ferroelectrics show prominent characteristics such as their being light weight, flexible, and environmentally friendly. However, they have no remarkable applications because their Curie temperature is below room temperature, and their magnitude of the spontaneous polarization is smaller than that of inorganic ferroelectrics. Recently, the performance of low-dimensional molecular organic ferroelectrics has been improved [6]; it was reported that a series of hydrogen-bonded organic molecular crystals show room-temperature ferroelectricity with a large magnitude of spontaneous polarization. For example, a single-component molecular crystal of croconic acid (4,5-dihydroxy-4-cyclopentene-1,2,3-trione) [7] shows a large magnitude of spontaneous polarization ($\sim$20 $\mu$C/cm$^2$), comparable to that of BaTiO$_3$ ($\sim$ 25 $\mu$C/cm$^2$). Thus, room-temperature hydrogen-bonded organic ferroelectrics are candidates for use in various electronic and photonic devices. Indeed, the second-order nonlinear susceptibility of croconic acid was reported to be fairly large [8], i.e., $\sim$ 2.5 $\times$ 10$^{-6}$ esu at 1000 nm, which is comparable to or higher than those of LiNbO$_3$ (1.2 $\times$ 10$^{-7}$ esu at 10640 nm) and dimethylamino-4-$N$-methylstilbazolium tosylate (DAST) (4.8 $\times$ 10$^{-6}$ esu at 1318 nm). Furthermore, we have recently found that a terahertz electromagnetic wave is emitted from a single crystal of croconic acid at room temperature [9]; the magnitude of the terahertz wave was comparable to that emitted from a typical terahertz wave emitter ZnTe.

Recently, a single-component molecular crystal, 2-phenylmalondialdehyde (PhMDA), has also been found to show ferroelectricity at room temperature [10]. It is composed of a molecule with the $\beta$-diketone enol O=C-C=C-OH group, as illustrated in Fig. 1(a). Hydrogen bonding (dotted line) connects molecules either along the [102] or $[\overline{1} 02]$  direction. The asymmetric off-center proton configurations break the space inversion symmetry of the crystal (space group of $Pna2_1$) along these directions. Thus, the vector composition in these directions would be the direction of the electric polarization of the crystal, i.e., $c$-axis. Indeed, the spontaneous polarization of $\sim$ 9 $\mu$C/cm$^2$ is induced along the $c$-axis [10]. Furthermore, it was reported that the Curie temperature of PhMDA exceeds 363 K [10]. 

In this paper, we report on the observation of room-temperature terahertz radiation from PhMDA by the irradiation of femtosecond laser pulses. By measuring the azimuth angle and laser power dependences, we found that the terahertz radiation from PhMDA is due to optical rectification via the second-order nonlinear optical effect. We also revealed that the magnitude of the terahertz radiation from PhMDA is large, comparable to that from ZnTe. This is attributable to the long coherence length in the range of 130--800 $\mu$m for the terahertz radiation from PhMDA. On the basis of the systematic optical measurements in the terahertz and visible frequency regions, we also examined the phase-matching condition for the terahertz radiation from PhMDA.

\section{Tensor components of second-order nonlinear optical susceptibility}

First, we explain a general mechanism of terahertz radiation in non-centrosymmetric media and describe the manifestation of the second-order nonlinear optical susceptibility of PhMDA. The second-order nonlinear polarization $P_{\rm NL}$ of non-centrosymmetric media can be expressed as $P_{\rm NL}=\epsilon_0\chi^{(2)}(\omega-\omega=0)E^\omega E^\omega$, where $\epsilon_0$ is the dielectric constant in vacuum, $E^\omega$ is the electric field of incident light, and $\chi^{(2)}$ is the second-order nonlinear optical susceptibility [11]. When the incident light is a femtosecond laser pulse with a finite spectral width of $\sim$ 10 THz, $P_{\rm NL}$ can be induced by difference frequency mixing of spectrum components, resulting in the emission of a terahertz wave. This process, called optical rectification in general, was recognized as the terahertz radiation mechanism in various non-centrosymmetric materials such as ZnTe, LiNbO$_3$ [12,13], DAST [14], and croconic acid [9]. 

In the electric dipole approximation, the non-zero tensor components of $\chi^{(2)}$ of PhMDA are $\chi^{(2)}_{xxz}$, $\chi^{(2)}_{xzx}$, $\chi^{(2)}_{yyz}$, $\chi^{(2)}_{yzy}$, $\chi^{(2)}_{zx}$, $\chi^{(2)}_{zyy}$, and $\chi^{(2)}_{zzz}$ [11]. Here, the $x$-, $y$-, and $z$-axes correspond to the crystallographic principal axes, $a$, $b$, and $c$, respectively. $P_{\rm NL}$ of PhMDA using the contradicted $d$ tensor can be expressed as 
\begin{eqnarray}
\begin{array}{ccc}
P_{\rm NL}=\epsilon_0\times\\
\left[
\begin{array}{cccccc}
0 & 0 & 0 & 0 & d_{15} &0 \\
0 & 0 & 0 & d_{24} & 0 &0 \\
d_{31} & d_{32} & d_{33} & 0 & 0 &0 \\
\end{array}
\right]
\left[
\begin{array}{c}
E_x^2\\
E_y^2\\
E_z^2\\
2E_yE_z\\
2E_zE_x\\
2E_xE_y\\
\end{array}
\right],
\end{array}
\end{eqnarray}
Here, $E_x$, $E_y$, and $E_z$ represent $x$-, $y$-, and $z$-components of the electric field of femtosecond laser pulses, respectively.

\section{Experimental procedure}
Single crystals of PhMDA were grown by the previously reported method (Ref. 10). We used single crystals with large $ac$-surfaces, which were obtained by cleaving the as-grown crystals. The crystal orientation was examined on a back Laue photograph. 

In the visible frequency region, we measured the polarized reflectance and transmittance spectra using a grating monochromator. 
In the terahertz frequency region, terahertz time-domain spectroscopy was performed in transmission geometry [see, Fig. 2(a)]. We used a photoswitching device prepared on a low-temperature-grown GaAs (LT-GaAs) [5] as the detector. The terahertz emitter was a 1-mm-thick (110)-oriented ZnTe crystal. 

For terahertz radiation experiments, we adopted a standard terahertz radiation detection setup using an LT-GaAs detector [5]. We used the laboratory coordinates as the $X$- and $Z$-axes [see, Fig. 3(a)]. The angle $\theta$ was defined as the angle of the $c$-axis of the crystal relative to the $X$-axis. Femtosecond laser pulses delivered from a mode-locked Ti:sapphire laser pulse (wavelength, 800 nm; repetition rate, 80 MHz; pulse width, 100 fs) were irradiated on the sample with a focused diameter of $\sim$ 25 $\mu$m. The laser power was fixed to 30 mW unless otherwise stated. The electric field $E_0$ of femtosecond laser pulses was set parallel to the $X$-axis. We used a wire-grid polarizer to detect the $X$-axis component of the radiated terahertz wave. 

All the measurements were performed at room temperature.

\section{Results and discussion}
\subsection{Polarized optical spectra in the visible frequency region}
The polarized reflectance $R$ and transmittance $T$ spectra in the visible frequency region are shown by red and blue lines in Fig. 1(b), respectively. We used a 810-$\mu$m-thick single crystal. The light electric field $E^\omega$ was set parallel to the $c$-axis, i.e., the direction of the spontaneous polarization. The increase in $R$ and the sharp drop of $T$ above 3 eV are attributed to the $\pi$-$\pi^*$ transitions [8] of the PhMDA molecule at approximately 3.9 eV. We performed the Kramers-Kronig transformation to extract the refractive index $n$ spectrum, which is shown by the red line in Fig. 1(c). It shows little dispersion below 3 eV. To estimate the optical group refractive index $n_{\rm g}$ spectrum, we fitted the $n$ spectrum using the Sellmeier relationship, which is expressed as
\begin{equation}
n=\sqrt{1+\frac{S_0\lambda_0^2}{1-(\lambda_0/\lambda)^2}},
\end{equation}
where $\lambda$ is the wavelength. The fitting result is indicated by the broken line. Using the obtained parameters $S_0 = 5.910$ m$^{-2}$ and $\lambda_0$ = 229 nm, we calculated the $n_{\rm g} (=| n-\lambda(dn/d\lambda)|)$ spectrum, which is shown by the green line in Fig. 1(c). $n_{\rm g}$ at 1.5 eV [the photon energy of femtosecond laser pulses used for terahertz radiation experiments] was estimated to be $\sim$ 1.86 [the vertical broken line in Fig. 1(c)]. We also show in Fig. 1(c) the absorption coefficient $\alpha$ spectrum derived from the measured $R$ and $T$ spectra. PhMDA has a transparent window in the energy range of 1--3 eV. For the photon energy of 1.5 eV, the penetration depth was evaluated to be $\sim$ 1.4 cm.

\subsection{Polarized optical spectra in the terahertz frequency region}

Next, we show the polarized $T$ spectrum for $E^\omega\parallel c$ in the terahertz frequency region. The schematic of the terahertz time-domain spectroscopy is shown in Fig. 2(a). We used a 810-$\mu$m-thick single crystal. The single-cycle terahertz wave generated from a (110)-oriented ZnTe is shown by the gray line in Fig. 2(b). The blue line indicates the terahertz wave passing through the sample. Owing to the increase in optical path length induced by the sample, the phase of the transmitted terahertz wave is delayed. In addition, the amplitude of the transmitted terahertz wave attenuates owing to the absorption of the sample. Figure 2(c) shows the power $T$ spectrum. The sharp decrease in $T$ above $\sim$ 1.7 THz would be attributed to the vibration mode of the molecules. Here, we focus on the optical properties below 1.8 THz. We numerically calculated the $n$ and $\alpha$ spectra, which are shown by the blue and green lines in Fig. 2(d); our estimation procedure is detailed in Ref. 15. Below 1.5 THz, $n$ is nearly constant ($\sim$ 2.2--2.4). The horizontal broken line indicates the value of $n_{\rm g}$ at 800 nm (1.86) [see Fig. 1(c)]. $\alpha$ is negligibly small in the same frequency region.

\subsection{Terahertz radiation experiments}

For terahertz radiation experiments, we used a 250-$\mu$m-thick single crystal. Since the penetration depth was $\sim$ 1.4 cm at 1.5 eV [Fig. 1(c)], PhMDA is transparent with respect to the incident femtosecond laser pulses. Figure 3(a) shows a schematic illustration of our setup for terahertz radiation experiments. By irradiation of femtosecond laser pulses at $\theta = 0^\circ$ in normal incidence, we found that the terahertz wave radiates into free space. A typical waveform of the terahertz electric field is shown in Fig. 3(b). It consists mainly of a single-cycle pulse with a temporal width of $\sim$ 0.5 ps. Figure 3(c) shows the Fourier transformed spectrum of the radiated terahertz wave. The radiated terahertz pulse contains the frequency component up to $\sim$ 1.9 THz with a peak frequency of $\sim$ 1.1 THz. 

To determine the terahertz radiation mechanism in PhMDA, we measured the azimuth angle $\theta$ dependence of the terahertz electric field at 0 ps ($E_{\rm THz}$). In this experiment, the sample was rotated in the $ac$-plane, as illustrated in Fig. 3(a); $\theta = 0^\circ$ corresponds to $E_0 \parallel c$, while $\theta = 90^\circ$ corresponds to $E_0 \parallel a$. $E_{\rm THz}$ as a function of $\theta$ is shown by circles in Fig. 3(d). $E_{\rm THz}$ reaches its maximum amplitude at 0$^\circ$, while its sign is reversed by rotating the sample by 180$^\circ$. In addition, $E_{\rm THz}$ becomes zero at 90$^\circ$. When optical rectification discussed in Sect. 2 is dominant for the terahertz radiation from PhMDA, the observed $\theta$ dependence of $E_{\rm THz}$ should be proportional to $P_{\rm NL}$ in Eq. (1). In our experimental geometry [Fig. 3(a)], $d_{33}$  and $d_{15}$  persist. We confirmed that $d_{15}$  is negligibly small, compared with $d_{33}$. In addition, we only detected the $X$-axis component of $E_{\rm THz}$. Therefore, $E_{\rm THz}$ is expressed by
\begin{equation}
E_{\rm THz}=d_{33}\cos^3\theta E_0^2,
\end{equation}
The solid line obtained using Eq. (3) reproduces well the observed $\theta$ dependence of $E_{\rm THz}$. We also measured the laser power dependence of $E_{\rm THz}$ at $\theta = 0^\circ$, which is shown by circles in Fig. 3(e). $E_{\rm THz}$ linearly increases with laser power ($E_0^2$) and is reproduced well by the solid line obtained using Eq. (3). These results indicate that optical rectification is the origin of terahertz radiation from PhMDA.

\subsection{Terahertz radiation characteristics}

To clarify the terahertz radiation characteristics of PhMDA, the estimation of coherence length for terahertz radiation is necessary. Using the values of $n$ in the terahertz frequency region [Fig. 2(d)] and $n_{\rm g}$ at 800 nm [Fig. 1(c)], we estimated the coherence length $l_{\rm c}$  for terahertz radiation. $l_{\rm c}$  is expressed as [16]
\begin{equation}
l_{\rm c}=\frac{\pi c}{\omega|n_{\rm g}-n_{\rm THz}|}, 
\end{equation}
where $c$ is the velocity of light. $l_{\rm c}$  as a function of frequency is shown by the red line in Fig. 4(a). $l_{\rm c}$  at 1.8 THz was evaluated to be $\sim$ 130 $\mu$m. With lowering frequency, $l_{\rm c}$ monotonically increases and reaches $\sim$ 800 $\mu$m at 0.5 THz. $l_{\rm c}$ in PhMDA is long, comparable to that in croconic acid ($\sim$ 100--900 $\mu$m) [9]. We measured the terahertz wave radiated from a 1-mm-thick (110)-oriented ZnTe crystal under the same experimental condition; ZnTe is a commercially available terahertz wave emitter. The electric field of the terahertz wave of PhMDA is $\sim$ 4 times smaller than that of ZnTe. By taking into account the sample thickness, it should be corrected by a factor of 4. Thus, the intrinsic terahertz radiation efficiency of PhMDA is found to be comparable to that of ZnTe. This is attributable to a long $l_{\rm c}$ in PhMDA.

Next, we discuss the spectral shape of the radiated terahertz wave [Fig. 4(c)] by taking into account the detector characteristic and absorption of the sample in the terahertz frequency region. To include the effect of absorption, we calculated the effective generation length for the terahertz radiation $L_{\rm gen}$ [17], which is expressed by
\begin{equation}
L_{\rm gen}=\left[\frac{1+\exp(-\alpha d)-2\exp\left(-\frac{\alpha}{2} d \right)\cos\left(\frac{\omega}{c}\mid n_{\rm o}-n_{\rm g}\mid d\right)}{\left(\frac{\alpha}{2}\right)^2+\left(\frac{\omega}{c}\right)^2(n_{\rm o}-n_{\rm g})^2}\right]^{1/2}.
\end{equation}
where $d$ is the sample thickness. $L_{\rm gen}$  represents the length along the depth direction, i.e., the $b$-axis, for the region contributing to the terahertz radiation. $L_{\rm gen}$ as a function of frequency is shown by the green line in Fig. 4(a). Next, we estimate the instrumental function $H_{\rm inst}$ in our experimental setup by considering the frequency response function $H_{\rm res}$ of the LT-GaAs detector. The amplitude of $H_{\rm res}$ can be obtained by the Fourier transform of the convolution of $\sigma(t)$ and the time profile of the incident femtosecond laser pulse. Here, $\sigma(t)$ is the photogenerated conductivity of LT-GaAs, which is simply expressed as $\sigma(t) \propto \exp(-t/\tau_{\rm carrier})$ with the carrier lifetime $\tau_{\rm carrier}$ ($\sim$ 0.3 ps) [18]. We include the effect of diffraction on the terahertz wave at the detector position. Then, $H_{\rm inst}$ is expressed as $H_{\rm res} \times I_{\rm THz} \times \omega^2$ [15], where $I_{\rm THz}$ represents the power spectrum of the ideal terahertz wave induced by optical rectification. Figure 4(b) shows the frequency dependence of terahertz radiation efficiency derived from $H_{\rm inst} \times L_{\rm gen}^2$, which reproduces the measured power spectrum of the terahertz radiation shown in Fig. 4(c). This indicates that the spectral shape of the radiated terahertz wave is mainly determined by the detector response and the effect of absorption of the sample in the terahertz frequency region.

Finally, we discuss the high potential of PhMDA as a terahertz wave emitter on the basis of systematic optical measurements in the visible and terahertz frequency regions. As shown in Fig. 2(d), below 1.5 THz, $n$ is nearly constant ($\sim$ 2.2--2.4) and $\alpha$ is negligibly small. When the photon energy of incident femtosecond laser pulses was set to be 3 eV, the phase-matching ($n_{\rm g} \approx n$) condition can be satisfied [Fig. 1(c)]. Furthermore, $\alpha$ at 3 eV is very small. Thus, PhMDA is a highly potential candidate for a terahertz wave emitter.

\section{Summary}
In summary, we observed the terahertz radiation from 2-phenylmalondialdehyde (PhMDA) at room temperature by the irradiation of femtosecond laser pulses. On the basis of the measurements of the azimuth angle and laser power dependences, we concluded that the terahertz radiation mechanism is optical rectification via the second-order nonlinear effect. The observed magnitude of the terahertz wave in PhMDA is comparable to that of the commercially available typical terahertz wave emitter ZnTe. This arises from a long coherence length in the range of 130--800 $\mu$m for the terahertz radiation from PhMDA. We also showed that better phase-matching for terahertz radiation can be achieved by setting the photon energy of the incident femtosecond laser pulses to $\sim$ 3 eV. 

\begin{acknowledgment}
This work was supported in part by Grant-in-Aids from MEXT (Nos. 25247049, 25600072, and 25-3372).
\end{acknowledgment}

\clearpage
\begin{figure}[t]
\begin{center}
\includegraphics[width=0.45\textwidth]{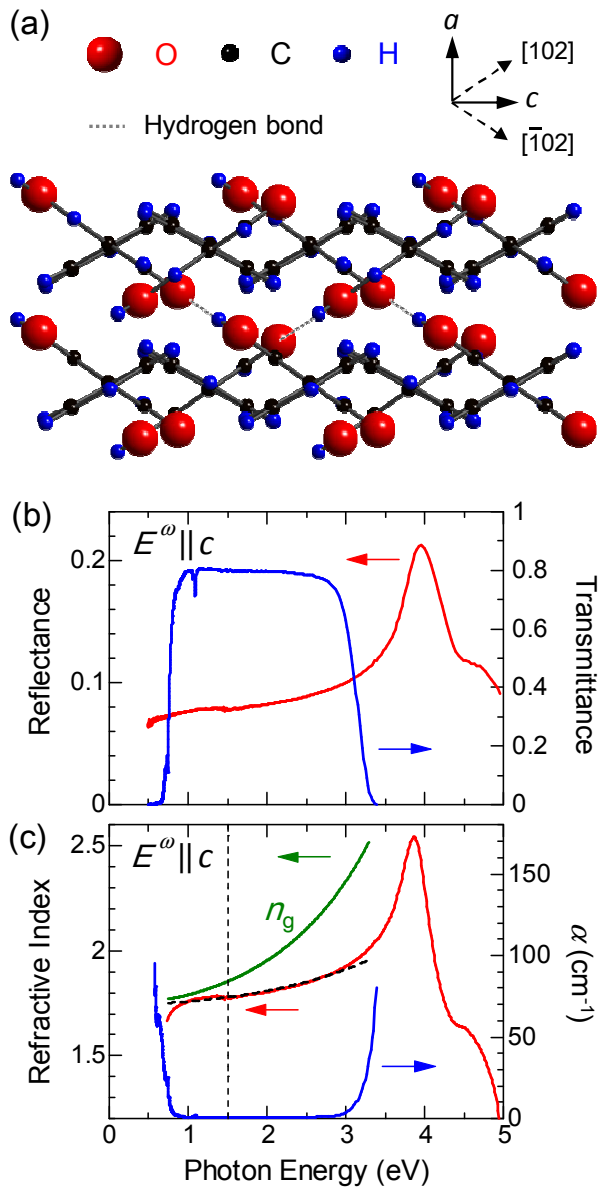}
\end{center}
\caption{(Color online) (a) Schematic illustration of the crystal structure of PhMDA. (b) Reflectance and transmittance spectra in $E^\omega\parallel c$, measured at room temperature. (c) Refractive index and absorption coefficient $\alpha$ spectra. The broken line indicates the fitting result using the Sellmeier relationship. We obtained the optical group refractive index $n_{\rm g}$ spectrum. The photon energy (1.5 eV) of femtosecond laser pulses is indicated by the vertical broken line.}
\end{figure}

\clearpage

\begin{figure}[t]
\begin{center}
\includegraphics[width=0.45\textwidth]{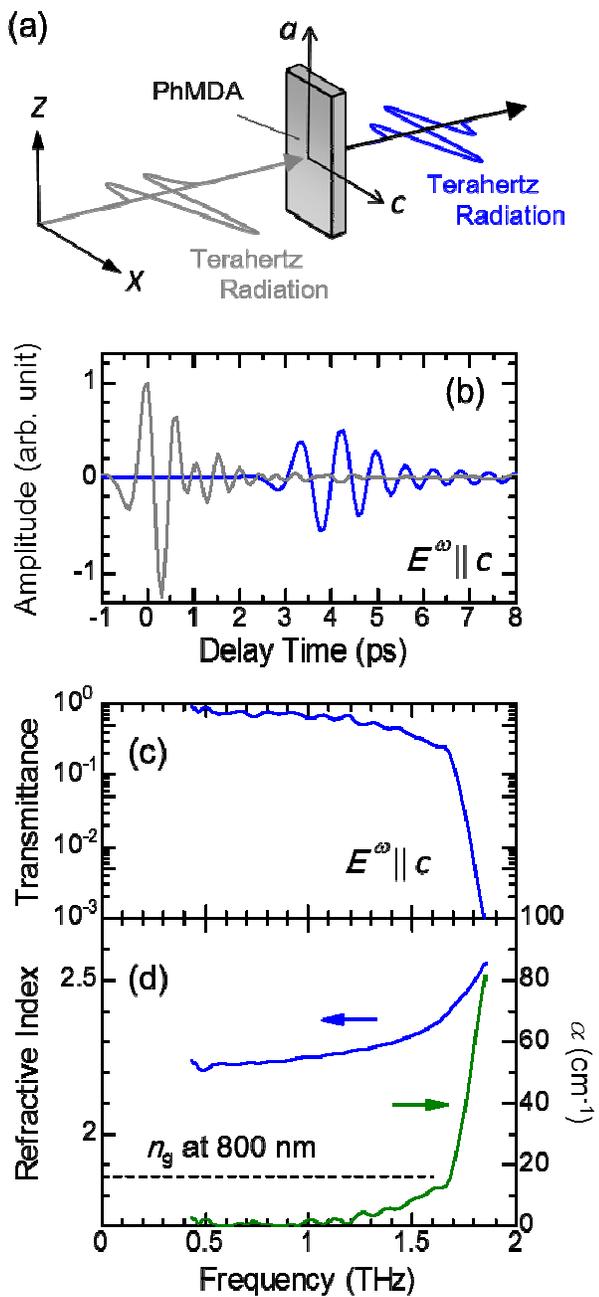}
\end{center}
\caption{(a) Schematic illustration of terahertz time-domain spectroscopy in transmission geometry. (b) Terahertz waveforms with and without sample. (c) Transmittance spectrum of PhMDA for $E^\omega\parallel c$, measured at room temperature. (d) Refractive index and absorption coefficient $\alpha$ spectra. }
\end{figure}

\clearpage

\begin{figure*}[t]
\begin{center}
\includegraphics[width=0.78\textwidth]{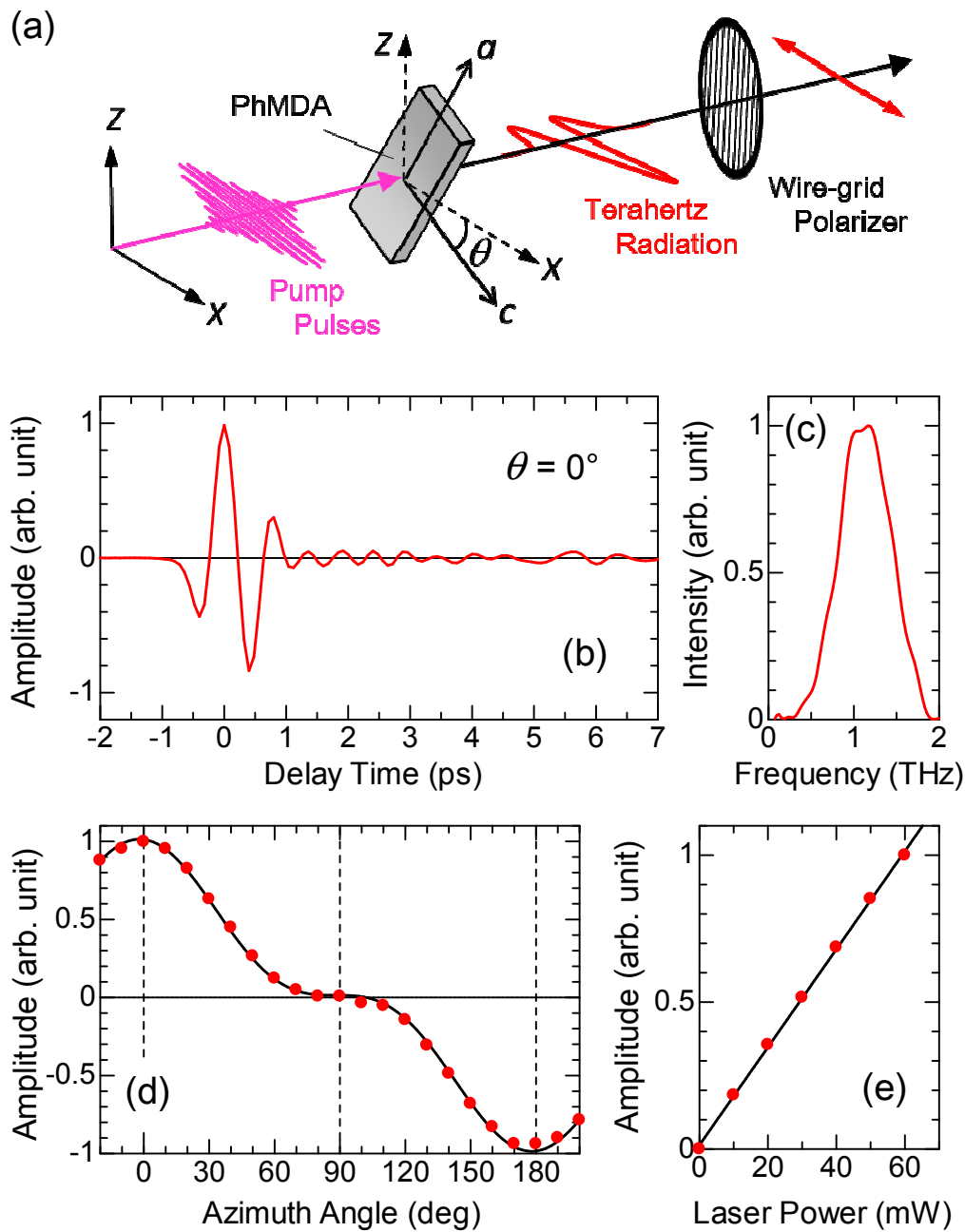}
\end{center}
\caption{(Color online) (a) Schematic illustration of the experimental setup for terahertz radiation. (b) Radiated terahertz waveform in PhMDA measured at room temperature. (c) Power spectrum of the waveform shown in (b). (d) Azimuth angle and (e) laser power dependences of the terahertz electric field at 0 ps. The solid lines in (d) and (e) are results of least-square fitting using Eq. (3).}
\end{figure*}

\clearpage

\begin{figure}[t]
\begin{center}
\includegraphics[width=0.38\textwidth]{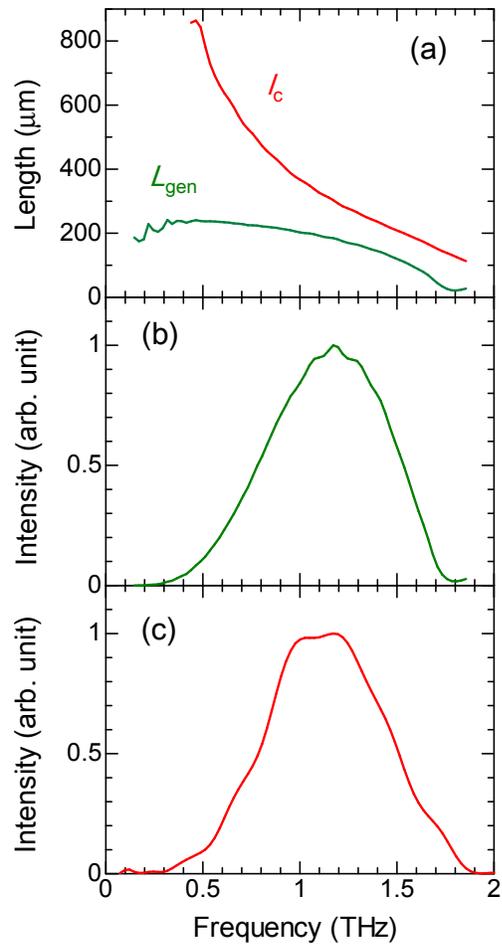}
\end{center}
\caption{(Color online) (a) Coherence length $l_{\rm c}$ and effective generation length $L_{\rm gen}$ for the terahertz radiation as a function of frequency. (b) Calculated and (c) measured power spectra of the generated terahertz wave. 
}
\end{figure}

\end{document}